\documentclass[aps,onecolumn,floatfix,notitlepage]{revtex4-1}
\usepackage{graphicx}
\usepackage{color}

\begin{document}

\title{Multiscale entanglement in ring polymers under spherical confinement}
 \author{Luca Tubiana$^1$, Enzo Orlandini$^2$, Cristian Micheletti$^{1}$} 

\affiliation{ \centerline{$^1$ SISSA - Via Bonomea 265 - I-34136, Trieste -
Italy}  \\
\centerline{$^2$ Dipartimento di Fisica ``G. Galilei'' and Sezione INFN -
Via Marzolo 8 - I-35100 Padova, Italy}}

\begin{abstract}
  The interplay of geometrical and topological entanglement in
  semiflexible knotted polymer rings confined inside a spherical cavity is
  investigated using advanced numerical methods. By using stringent
  and robust algorithms for locating knots, we characterize how the
  knot length $l_k$ depends on the ring contour length, $L_c$ and
  the radius of the confining sphere, $R_c$. In the no- and
  strong-confinement cases we observe weak knot localization and
  complete knot delocalization, respectively.  We show that the
  complex interplay of $l_k$, $L_c$ and $R_c$ that seamlessly bridges
  these two limits can be encompassed by a simple scaling argument
  based on deflection theory. The same argument is used to rationalize
  the multiscale character of the entanglement that emerges with
  increasing confinement.
\end{abstract}
 \maketitle If we tie a knot in a
piece of rope and pull the latter at its extremities, the knotted part
will become readily distinguishable from the rest of the rope because
it localizes. Indeed, for a very long rope only a neglible fraction of
its contour length will be required to accommodate the
knot~\cite{Buck:1998:Nature,*Katritch:2000:PRE,*Millett:2005:Macromol}.
In this intuitive example of \emph{knot localization}
the topological (global) entanglement embodied by the knot does not
interfere with the geometrical (local) entanglement of the rest of the
rope.  This is hardly the case for polymers that
circularize in equilibrium. In such rings, knots are abundant~\cite{Sumners&Whittington:1988:J-Phys-A} and the
interplay between topological and geometrical entanglement impacts
significantly the molecules' dynamical, mechanical and metric
properties~\cite{Katritch_et_al:1996:Nature,Arai:1999:Nature,Quake:1994:Phys-Rev-Lett,*Moore:2004:PNAS}.
Understanding how and to what extent global and local entanglement are
related is a major open issue in polymer
physics~\cite{Orlandini&Whittington:2007:Rev-Mod_Phys,*PhysRepReview2011,*Kamenetskii_SPU_1981}
with ramifications in key biological
contexts\cite{0953-8984-22-28-283102}, especially those related to
genome organization in eukaryotes, bacteria and
viruses~\cite{Grosberg_EPL_1993, *Lieberman-Aiden09102009,*Rosa:2008:PLOS,*Grosberg:2011:inpress,Jun_and_Mulder:2006:PNAS,Arsuaga:2005:Proc-Natl-Acad-Sci-U-S-A:15958528,Marenduzzo:2009:Proc-Natl-Acad-Sci-U-S-A:20018693,Micheletti:2006:J-Chem-Phys:16483240}.

A first breakthrough in the problem could be made by
establishing what fraction of the polymer ring is
occupied by the knot(s) and whether  this
measure is sufficiently robust or depends instead on the  geometrical
entanglement degree.
As a prototypical context to examine this problem we consider
semi-flexible, self-avoiding rings of cylinders with the simplest
knotted topology, a $3_1$ (trefoil) knot, and subject to isotropic spatial
confinement. In such a system, by varying the size of the confining
region, the degree of geometrical complexity can be changed and related
to the equilibrium size of the knotted ring portion.

For definiteness, the rings properties are set to
  match dsDNA. Specifically, the cylinder diameter and long axis are
  set equal to $d=2.5nm$ and $b=10nm$, respectively. The latter
  quantity is ten times smaller than the DNA Kuhn length (equal to
  twice the persistence length, $l_p=50nm$) thus ensuring a fine
  discretization of the model DNA. The system energy includes steric
  hindrance of non consecutive cylinders plus a bending potential:
\begin{equation}
E_b = - K_bT \frac{l_p}{b}\sum_{i=1}^{N} \vec t_i\cdot \vec t_{i+1}
\end{equation}
\noindent where $\vec t_i$ is the orientation of the axis of the $i$th
cylinder, $\vec{t}_{N+1} \equiv \vec{t}_1$ and the temperature $T$ is set to 300K. We considered rings
of $N=50,...250$ cylinders,  corresponding to countour lengths, $L_c =
N b$ ranging from 500$nm$ to 2.5 $\mu$m. This range allows for probing
changes in knot localization going from semiflexible to fully-flexible rings~\cite{Alim_Frey_2007_PRL} as well
as examing the effect of the interplay between $L_c$, $l_p$ and the
radius of the confining sphere, $R_c$. For simplicity of notation in the
following all lengthscales are expressed in units of $b$.

Because compact ring configurations are entropically
  disfavoured with respect to unconstrained ones, simple stochastic
  sampling schemes cannot be effectively used to generate spatially
  confined rings~\cite{PhysRepReview2011}. An analogous entropic
  attrition works against having a sizeable population of knots of a
  given type, trefoils in our case, at all levels of
  confinement~\cite{PhysRepReview2011}. To overcome these two difficulties we used 
  a biased Multiple-Markov-chain sampling scheme~\cite{Micheletti:2006:J-Chem-Phys:16483240} 
   with 24 Markovian replicas.
  For each replica, ring configurations are evolved using crankshaft
  and hedgehog Monte Carlo
  moves~\cite{klenin_1988_crankshaft_hedgehog}. The moves preserve the
  length of the rings, but not the topology, consistently with ergodicity
  requirements~\cite{Millett_Calvo_ergodicity,PhysRepReview2011}. In fact, even when the
  configurations before and after the move are self-avoiding, the move
  itself may involve self-crossings of the chain. A newly generated
  ring, $\Gamma$, is accepted according to the standard Metropolis
  criterion 
  with canonical weight, $\exp{[-U(\Gamma)/K_B T]}$, where the 
  potential energy, $U$ is suitably chosen to enhance the
  populations of compacts trefoil rings.  Specifically, if $\Gamma$ is
  non self-avoiding or has a too complicated topology (8 or more
  minimal crossings according to the Alexander polynomial
  \cite{Kamenetskii_SPU_1981}) then $U$ is set equal to infinity, otherwise
  $U(\Gamma)=E_b(\Gamma) + p\, R_c(\Gamma)$. In the latter expression,
  $R_c$ is the radius of the sphere containing $\Gamma$ (calculated as
  the largest distance of the cylinder vertices from the ring
  geometric center) and $p$ is a confining field whose value is set
  differently for the various replicas so to sample rings with different
  degree of confinement.

For each value of $N$ this procedure was used to
  generate at least $\sim10^7$ configurations with varying compactness
  and topology. For the {\em a posteriori} data analysis we picked,
  out of the generated rings, an uncorrelated subset with $3_1$
  topology (positively established with the KNOTFIND
  algorithm~\cite{Knotscape}). Finally a thermodynamic reweighting
  technique~\cite{Micheletti:2006:J-Chem-Phys:16483240} is used to
  remove the $p$ bias and obtain canonical averages for the observables.

\begin{figure}[t]
\includegraphics[width=0.8\textwidth]{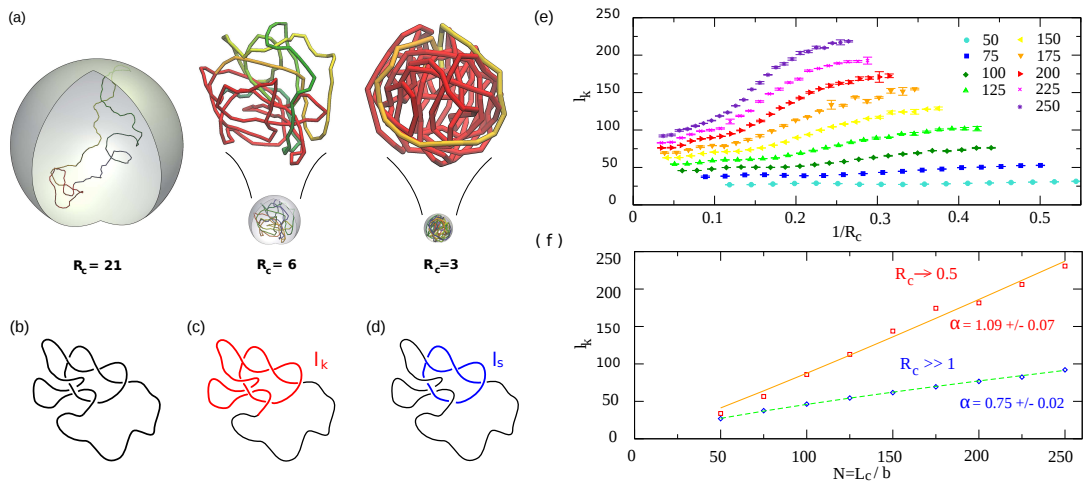}
\caption{ (a) Typical configurations of trefoil-knotted rings with $N  = 200$ cylinders at different levels of confinement. The encapsidated configurations for $R_c=6$ and $R_c=3$ are magnified for ease of visualization.
The knotted portion of the rings is shown in
  red. (b) Sketch of a trefoil-knotted ring; its knotted portion is highlighted
  in red in panel (c) and its shortest knotted arc is shown in blue in panel (d).
  (e)  Average knot length, $l_k$, as a function of $1/R_c$ for rings of varying number of cylinders, $N=50...250$. 
  (f) Dependence of $l_k$ on $L_c$ in the  no- and strong-confinement limits; the lines represent the best-fitting power laws. The strong-confinement data correspond to the limiting values approached exponentially by the curves in panel (e). The exponential fit was limited to the data points for which $l_k > 0.6 L_c$.}
\label{fig:def_cmp}
\end{figure}

Figure~\ref{fig:def_cmp}a illustrates both the range of spherical
confinement spanned by the sampled rings as well as
  the arclength covered by the knot. As a first method to identify the
  knotted portion we look for the arc, $\Gamma_1$, that (i) has the
$3_1$ topology while the complementary subchain, $\Gamma_2 \equiv
\Gamma \setminus \Gamma_1$ is unknotted; (ii) cannot be further
shortened without losing the full-ring $3_1$ topology and finally
(iii) can be extended continuously to encompass the whole ring
retaining the $3_1$ topology and the unknottedness of the
complementary arc, $\Gamma_2$. The topological state
  of the open arcs $\Gamma_1$ and $\Gamma_2$ is established after
  bridging their termini with the minimally-interfering closure of
  ref.~\cite{Tubiana:2011:PTP}. 

The above scheme for locating the knot
  can be seamlessly and robustly applied at all levels of ring
  compactification. As such, it considerably extends the scope of
  previous investigations where, depending on the free or collapsed
  nature of the rings, the knotted-arc identification had to be made
  with different and inequivalent methods, including ones affecting
  chain geometry by a rectification procedure or by external
  constraints such as a sliplink\cite{Marcone:2007:PRE,Virnau:2005:J_Am_Chem_Soc}.

Two important facts emerge from Fig.~\ref{fig:def_cmp}e: for a given
ring contour length, $L_c$, the increased confinement is parallelled
by an increase of the length of the knotted arc, $l_k$; at the same
time, for a given size of the confining sphere one has that $l_k$
increases with $L_c$.  As a first step towards a comprehensive
rationalization of the results we examine the $l_k$ versus $L_c$
behaviour in the no- and strong-confinement limits, which are shown in
Fig.~\ref{fig:def_cmp}f.

In the unconstrained case, it is seen that $l_k$ increases as a power of
$L_c$: $l_k(R_c \to \infty) \propto L_c^{\alpha}$ with
\textcolor{black}{$\alpha=0.75 \pm 0.02$.} The sublinear increase implies that $l_k/L_c$ vanishes for
increasing $L_c$, resulting in a weak knot localization. This
is consistent with previous results on knot localization for 
unconstrained rings both on- and off-lattice where, however, 
different values of $\alpha$ ranging from $0.54$ to $0.75$ were
reported~\cite{Marcone:2007:PRE,Virnau:2005:J_Am_Chem_Soc,*Mansfield&Douglas:2010:J-Chem-Phys}.
It also agrees qualitatively with recent experimental findings on
knotted circular DNA adsorbed on mica surfaces based on single
molecule imaging techniques~\cite{Ercolini:2007:PRL}.

A power-law dependence of $l_k$ on $L_c$ 
  provides a good fit of the data for strong confinement too but in
  this case \textcolor{black}{$\alpha=1.09 \pm 0.07 $}. This is compatible with a
  linear dependence of $l_k$ on $L_c$ thus
  implying a full delocalization of the knot in equilibrated rings
  subject to severe three-dimensional confinement. To the best of our
  knowledge this fact has not been established before. However, it is
  worth pointing out that knot delocalization has been previously observed for $\theta$-collapsed knotted
  rings~\cite{Marcone:2007:PRE,Virnau:2005:J_Am_Chem_Soc,*Mansfield&Douglas:2010:J-Chem-Phys}. The
  analogy with our findings is noteworthy since the ring
  metric properties are dictated purely by equilibrium thermodynamics
  for collapsed rings while spatial constraints are also at play for
  confinement.

\begin{figure}[t]
\includegraphics[width=0.8\textwidth]{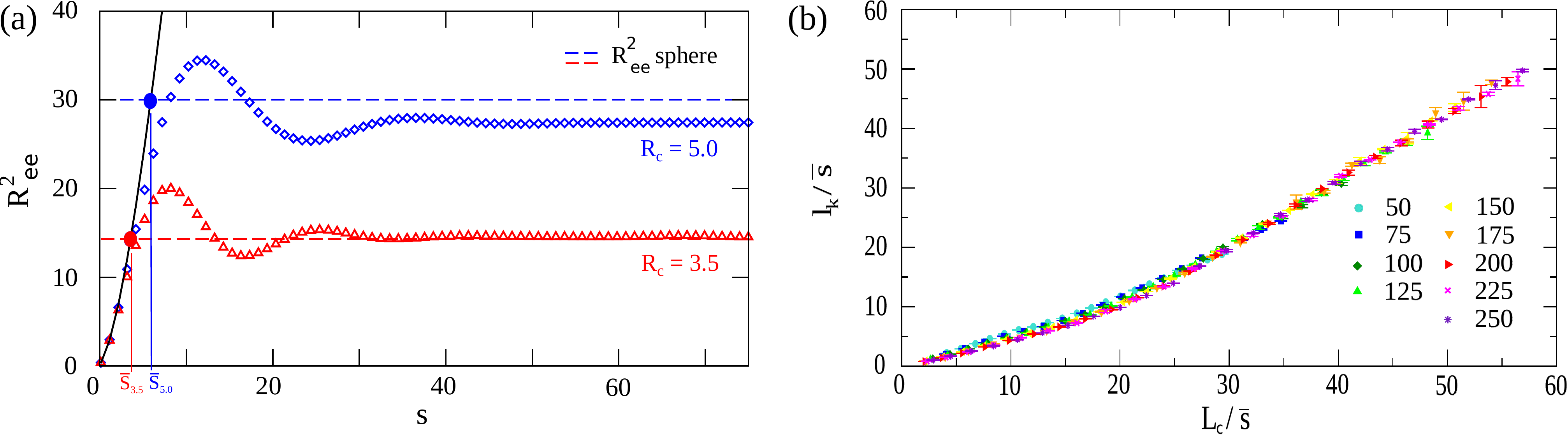}
\caption{ (a) Mean squared end-to-end distance, $R_{ee}^2$, of arcs of
  length $s$ in rings of $N = 150$ cylinders at different levels of
  confinement. The dashed horizontal line indicates the reference
  value equal to $6/5R_c^2$ while the continuous line shows the
  expected $R_{ee}^2$ for an unconstrained, non-self-avoiding Kratky-Porod (KP) chain. The
  intersection of these reference lines determines the nominal deflection
  arclength, $\bar{s}$ highlighted by the dropline. (b)
  Collapse of the same data points as in Fig.~\ref{fig:def_cmp}e obtained by
  rescaling $l_k$ and $L_c$ by $\bar{s}$. \textcolor{black}{The slope of the line
  fitting the data for ${L_c \over \bar{s}} > 40$ is $1.09\pm
  0.02$.} } \label{fig:gyr}
\end{figure}

We next turn our attention to the intermediate regime
  intervening between the no- and strong-confinement limits and show
  that the complex behaviour of Fig.~\ref{fig:def_cmp}e can be rationalized in terms
  of a surprisingly simple relationships involving the length scales $L_c$, $R_c$ and $l_k$.

To this purpose we first observe that upon increasing confinement the
geometric characteristics of the rings resemble those of
uniformly filled spheres. In particular the eigenvalues of the tensor
of gyration of rings become close to the eigenvalue of a
uniformly-filled sphere (see supporting material).

The approximately uniform filling of the confining sphere by mildly- or
strongly-confined rings is well illustrated in Fig.~\ref{fig:gyr}a
which portrays the mean squared end-to-end separation, $R_{ee}^2(s)$
of ring portions of different arclengths, $s$. In fact, the profile of
$R_{ee}^2(s)$ is noticeably flattened and its plateau value is about
$6/5 R_c^2$, which is the mean square distance of two points
inside a spherical volume of radius $R_c$. 
The value of $s$ at which the plateau value of $R_{ee}^2$ is reached
corresponds to the typical contour length, $\bar{s}$, of a discrete worm-like
(Kratky-Porod) chain with mean squared end-to-end distance equal to $6/5
R_c^2$ 
\begin{equation}
R_{ee}^2(\bar{s}) = 2l_p {\bar s} \left [1 - \frac {l_p}{{\bar s}} \left (1 - e^{-\frac {\bar{s}} {l_p}} \right) \right] = \frac{6}{5} R_c^2 \ .
\label{eqn:brutta}
\end{equation}
\noindent $\bar{s}$ is the nominal arclength of unrestricted ring-portions that are
just long enough to hit the sphere boundary and, by inverting
(\ref{eqn:brutta}), can be expressed in terms of the principal branch of the
Lambert W function~\cite{Knuth_lambert_w_funct}, 
\begin{equation}
\bar{s}/ l_p = 1 + y + W(-\exp[ -1 -y])
\end{equation}
\noindent where $y \equiv 3 R_c^2 / 5 l_p^2$. We shall refer
to $\bar{s}$ as the deflection arclength, in analogy with
the deflection argument originally introduced for polymers
in channels \cite{Odijik:1983:Macromolecules} and later
applied to other contexts, such as crumpled
globules~\cite{Grosberg_EPL_1993}.

The deflection arclength is  crucial
for rationalizing the results of Fig.~\ref{fig:def_cmp}e and
~\ref{fig:def_cmp}f.
In particular it is plausible that confined rings that
  experience the same nominal number of deflections at the spherical
  hull boundary should have statistically similar knot lengths once
  rescaled by $\bar{s}$.
This conjecture is supported by the scatter plot of
Fig.~\ref{fig:gyr}b where the rescaled knot length is plotted against
the number of nominal deflections $L_c/\bar{s}$ 
for rings of all considered lengths, $50 \le N \le
250$, and degrees of confinement.
Despite the heterogeneity of the original data sets (see Fig.~\ref{fig:def_cmp}e), the plot displays a striking collapse of the data
points.  A broadening of the curves is seen only in the limit of
unconstrained rings, where indeed, the deflection length looses
meaning.  We stress that, because $\bar{s}$ is calculated
deterministically, no single adjustable parameter was introduced to
obtain the collapse in Fig.~\ref{fig:gyr}b. 
\begin{figure}[t]
\includegraphics[width=0.8\textwidth]{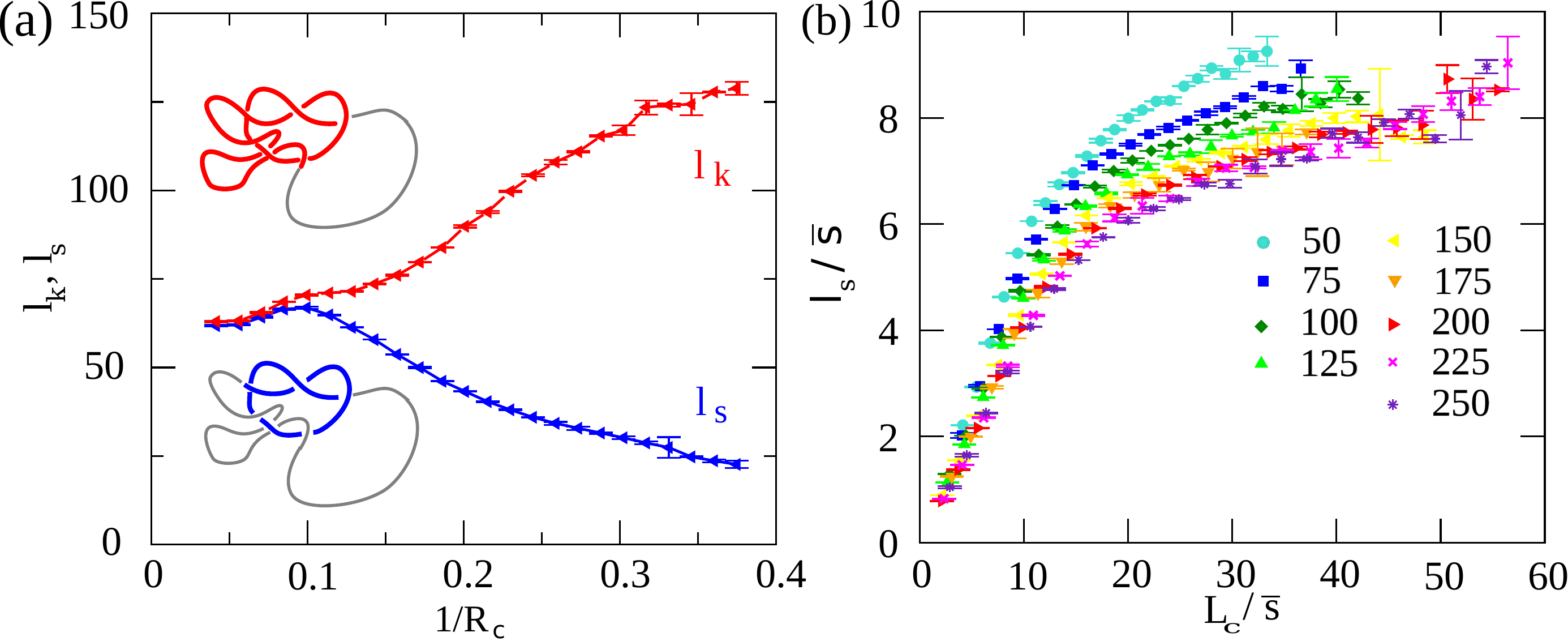}
\caption{ (a) Dependence of knot length, $l_k$, and the shortest knot
  length, $l_s$, on confinement for rings of $N = 150$ cylinders.
  (b) Scatter plot of $l_s/\bar{s}$ versus $L_c/\bar{s}$ for rings of
  all lengths and degree of confinement.}
\label{fig:both_schemes}
\end{figure}
The collapsed curve exhibits an asymptotic linear trend thus
reinforcing the delocalization result of Fig.~\ref{fig:def_cmp}f.

We conclude the study of the interplay of ring geometry and
topology by showing that, by increasing the ring compactification, a
non-trivial type and degree of entanglement is observed at
progressively smaller arclengths.

To do so we search for the smallest arcs with $3_1$
topology.  As illustrated in Fig.~\ref{fig:def_cmp}c,d  such minimally-long knotted
  portion may differ from the proper knotted arc introduced before because
it does not necessarily satisfy criterion (iii). Consequently, it may
correspond to an ephemeral knot (because it can be contained in a
longer arc with different topology, e.g. an
unknot)~\cite{Millett_2009_JKTR}.

Fig.~\ref{fig:both_schemes}a illustrates how the length of the
shortest knotted arc, $l_s$, behaves for increasing confinement and
compares it with $l_k$. It is seen that for no- or midly-constrained
rings, when the ring geometrical entanglement is minimal, the two
measures are in good accord. However, upon increasing confinement they
progressively diverge. Notice that $l_s$ and $l_k$ provide lower and
upper limits to the range of arclengths over which non-trivial
topological entanglements (after arc closure) are
observed. Consequently, the divergence of the two metrics indicates
that the entanglement cannot be described by a single length scale but
it displays a multiscale behaviour that amplifies upon increasing
confinement.
Similarly to the
case of the proper knot length, $l_k$, the concept of deflection
arclength is useful for rationalizing the interplay of $l_s$, $R_c$ and
$L_c$. In fact, $l_s$ expectedly results from the balance
between (a) the increasing probability that a knot can be tied with a
minimal (nominal) number of deflected segments/cylinders for longer and more confined
rings and (b) the decreasing probability that the complementary arc is
unknotted.

In the plot of Fig.~\ref{fig:both_schemes}b it is seen that, for
increasing $L_c$, the $l_s/\bar{s}$ data tend to approach a limiting
curve that has a linear dependence on $L_c/\bar{s}$. The results
indicate that for medium and strong confinement the length of the
smallest, and possibly ephemeral, knot $l_s$ increases approximately
linearly both with the $L_c$ at fixed $R_c$ and with $R_c$ at fixed
$L_c$ (more data in supp. mat.).  Together with the behaviour of
$l_k$, the results provide a quantitative basis for
  the multiscale character of the ring entanglement that sets in for
  increasing confinement.

To summarize, we have used simulations to measure the length of the
topologically entangled portion ($l_k$) of knotted
self-avoiding ring polymers subjected to different
degree of spherical confinement. We find that for weak confinement,
when the geometrical entanglement of the chain is moderate, the knot
is weakly localized while for strongly packed rings it delocalizes
completely.  The delocalization trend at increasing confinement and
different $L_c$ follows the neat scaling behavior of
Fig.~\ref{fig:def_cmp}f which is rationalised in terms of the
deflection arclength.  The latter therefore appears to be a key
quantity to characterize the complex interplay of the geometry and
topology in confined polymer rings, which reverberates in the above
mentioned multiscale entanglement. This observation prompts the
question of whether an analogous concept and scaling argument can be
introduced to characterize and rationalize the entanglement found in
other systems of densely-packed circular polymers, such as ring melts
or theta-collapsed
rings~\cite{Grosberg_EPL_1993,Virnau:2005:J_Am_Chem_Soc,Marcone:2007:PRE}.
We wish to point out that our results have direct bearings on the
relevant problem of viral DNA ejection into infected
cells~\cite{Matthews:2009:Phys-Rev-Lett:19257792}.  In fact this
crucial step of the viral infection is not hindered by the presence,
in the packaged genome, of knots because the latter are highly
delocalised \cite{0953-8984-22-28-283102,Marenduzzo:2009:Proc-Natl-Acad-Sci-U-S-A:20018693}.

 We thank D. Marenduzzo and A. McGown for useful discussions. We acknowledge support from Italian Ministry of Education.

\newpage
\section*{Supporting material}

\subsection{Size and shape of confined rings}
\begin{figure}[h!]
\includegraphics[width=0.6\textwidth]{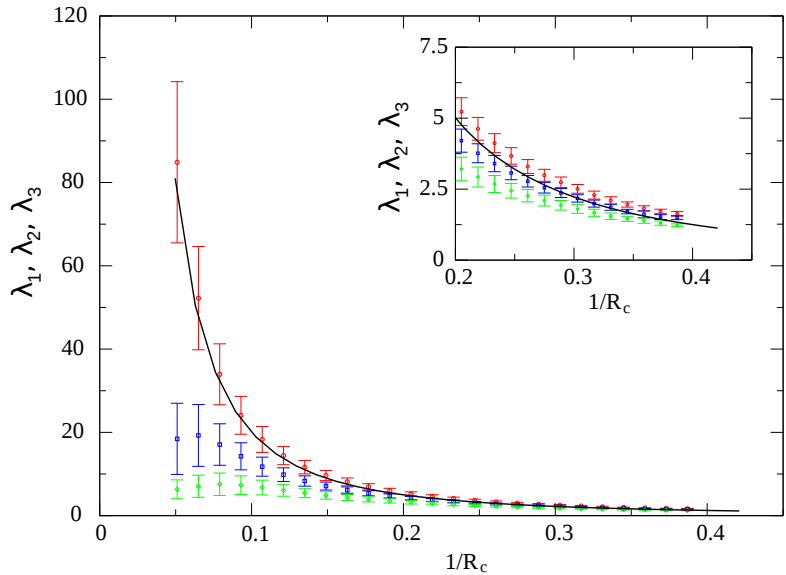}
\caption{The average eigenvalues of the rings gyration matrix,
  $\lambda_1$, $\lambda_2$, $\lambda_3$ (largest to smallest) are
  shown as a function of the radius of the confining spherical cavity,
  $R_c$. Errorbars report the standard deviation around the average
  value.  The inset shows the data for high spatial confinement.
  Going from no- to mild confinement, the average length of the rings
  principal gyration axis is substantially reduced while
  the other two are much less affected.  For example, the initial
  five-fold reduction of $\lambda_1$ in rings of $N=150$ cylinders is
  accompanied by a change of $\lambda_2$ of about $50\%$ and a 
  change of $\lambda_3$ not larger than 10\%. 
  The latter variation is non-monotonic, indicating a swelling
  of the mildly confined semi-flexible rings in the plane
  perpendicular to the (squeezed) primary gyration axis. Importantly,
  the strong shape anisotropy of unconstrained rings is largely lost
  even for mild confinement: already at $R_c=5$ all average gyration
  are compatible with the gyration eigenvalue of the {\em
    uniformly-filled } confining hull within the estimated errors.}
\end{figure}
\newpage
\subsection{Multiscale entanglement: detailed comparison of $l_k$ and $l_s$}

\begin{figure}[h!]
\includegraphics[width=0.6\textwidth]{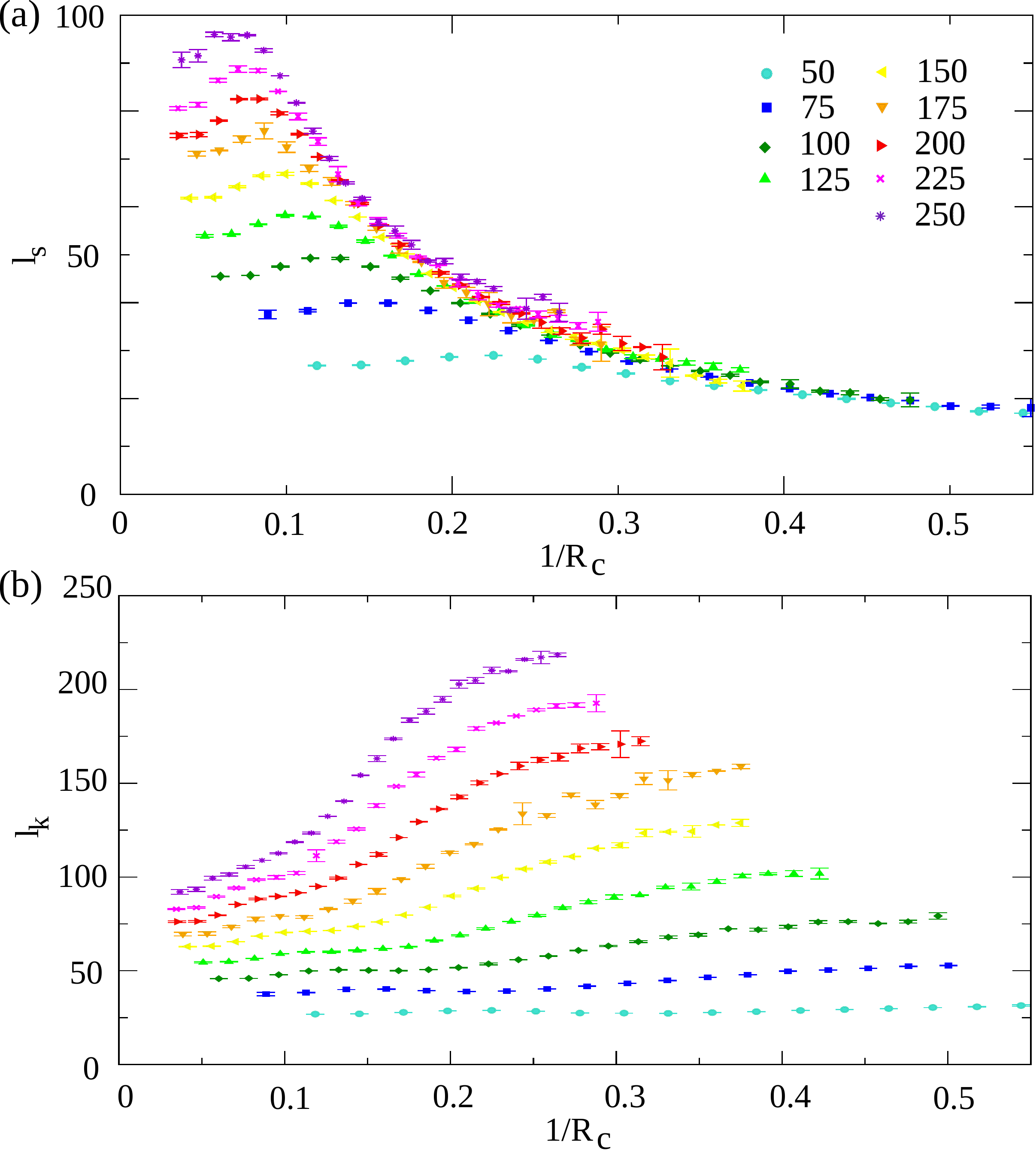}
\caption{Average length of shortest knotted arc (a) for
  rings of different contour lengths $L_c$ and increasing degree of
  confinement $1/R_c$. Notice that, at a given contour length, $l_s$
  becomes smaller for increasing confinement. Also, $l_s$ increases
  with $L_c$ at fixed degree of confinement. The average length of
  knotted portion, $l_k$, shown in Fig. 1 of the manuscript, is
  reported in panel (b) for ease of comparison.}
\end{figure}
\begin{figure}[h!]
\includegraphics[width=0.6\textwidth]{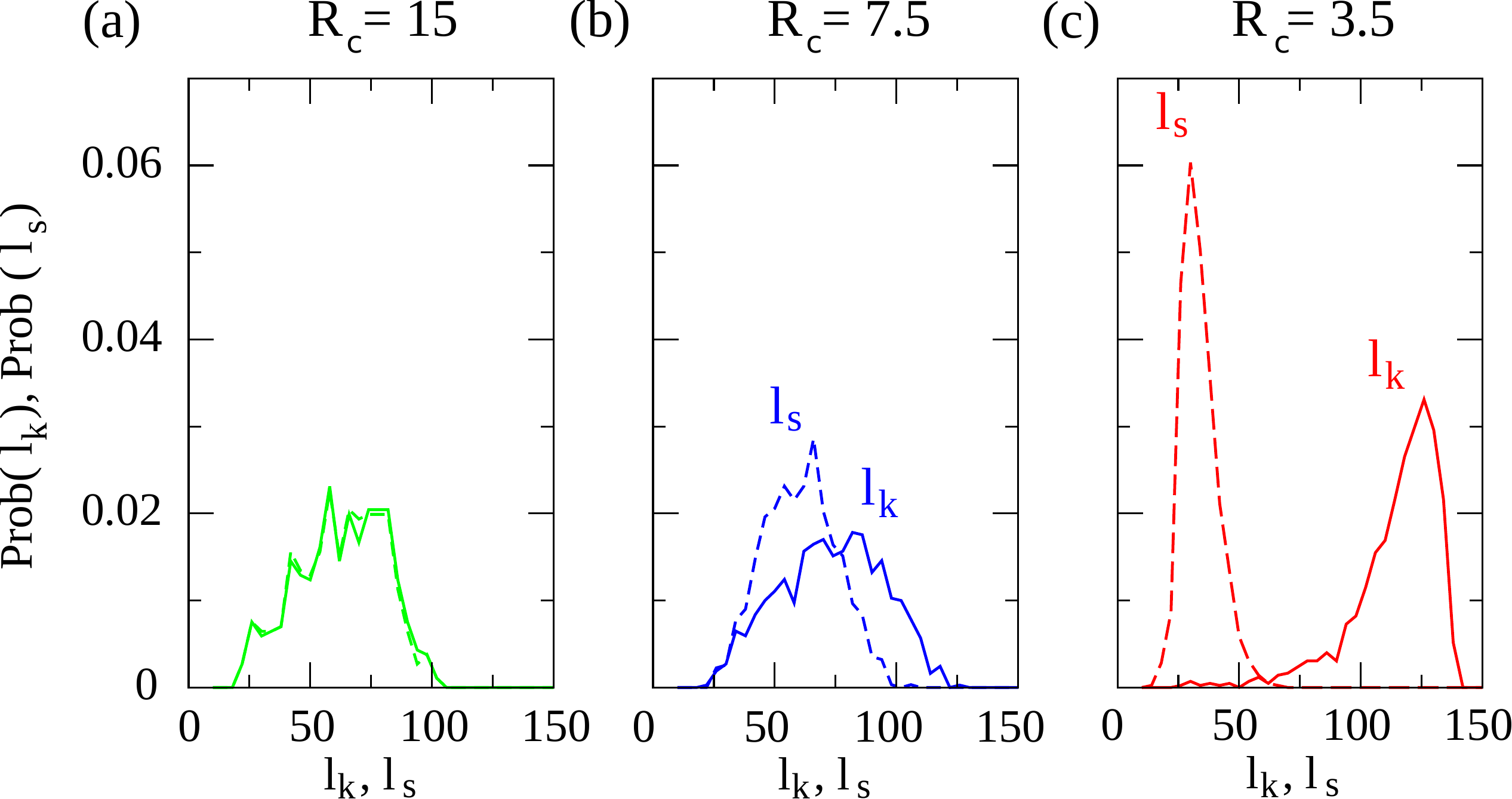}
\caption{Probability distribution of the lengths of the shortest
  knotted arc, $l_s$, and knotted portion, $l_k$, at different
  levels of confinement for rings of $N = 150$ cylinders. At no
  confinement the two distribution are pratically identical, see panel
  (a). However, their difference increases progressively with
  confinement, see panels (b) and (c).}
\end{figure}
\newpage

\subsection{Comparison with infinitely thin rings}

The robustness and generality of the multiscale character of the
entanglement found for self-avoiding rings is underscored by the fact
that an analogous phenomenology is found for confined infinitely-thin
trefoil-knotted rings, as reported in the following. The
infinitely-thin rings are Kratky-Porod chains: they have the same persistence lengths as the
self-avoiding ones discussed in the main paper, but have no excluded
volume.

Given the absence of steric hindrance, higher levels of
compactifications are achieved compared to self-avoiding rings.

\begin{figure}[h!]
\includegraphics[width=1.0\textwidth]{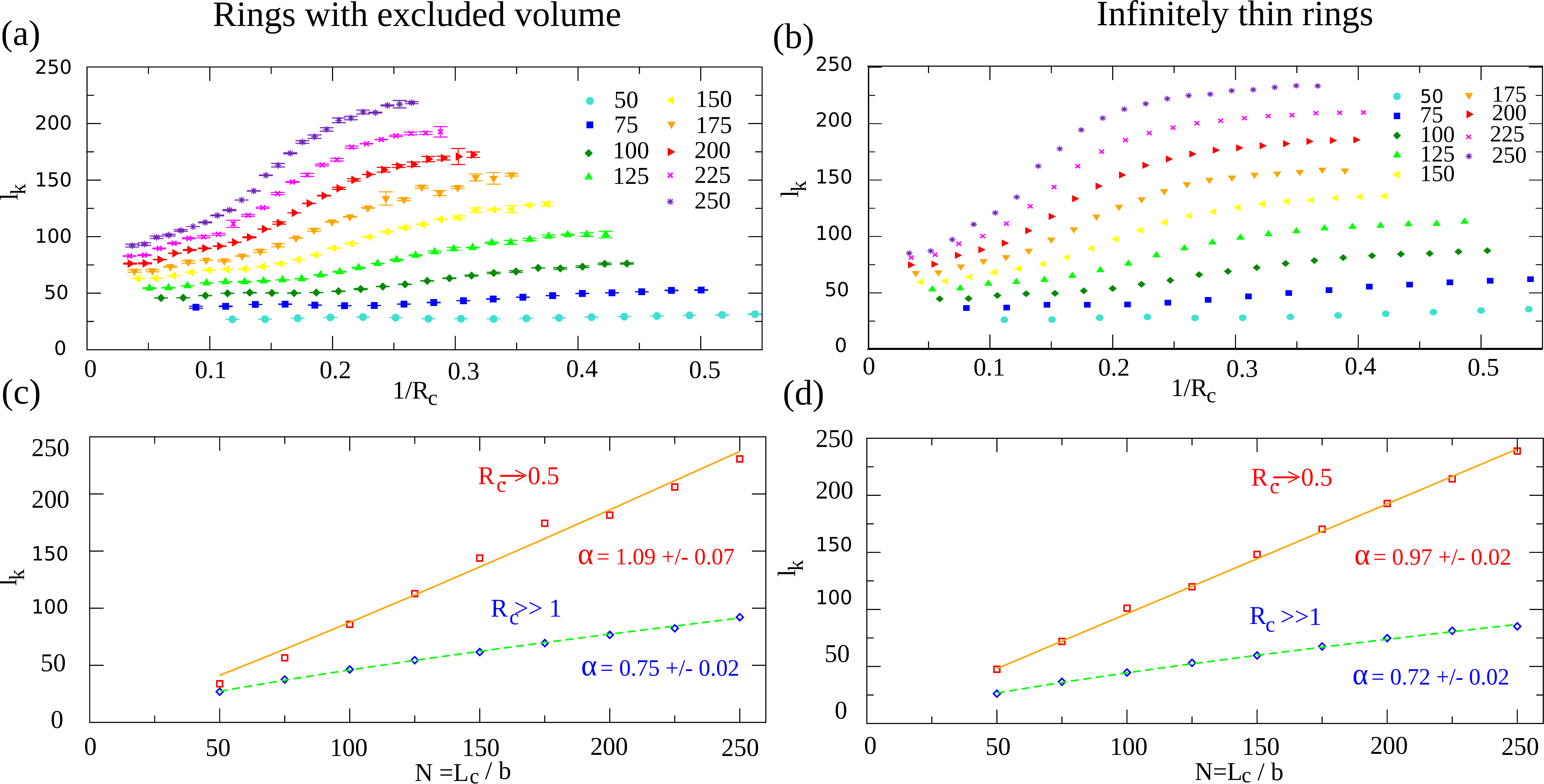}
\caption{(a) and (b): average knot length, $l_k$, as a function of
  $1/R_c$ for rings with and without excluded volume respectively. (c)
  and (d): Dependence of $l_k$ with $L_c$ in the no- and strong-
  confinement limits for rings with and without excluded volume
  respectively. Note that analogous qualitative behaviour of the two
  types of models.}\label{fig:supp_mat_4}
\end{figure}

\begin{figure}[h!]
\includegraphics[width=0.8\textwidth]{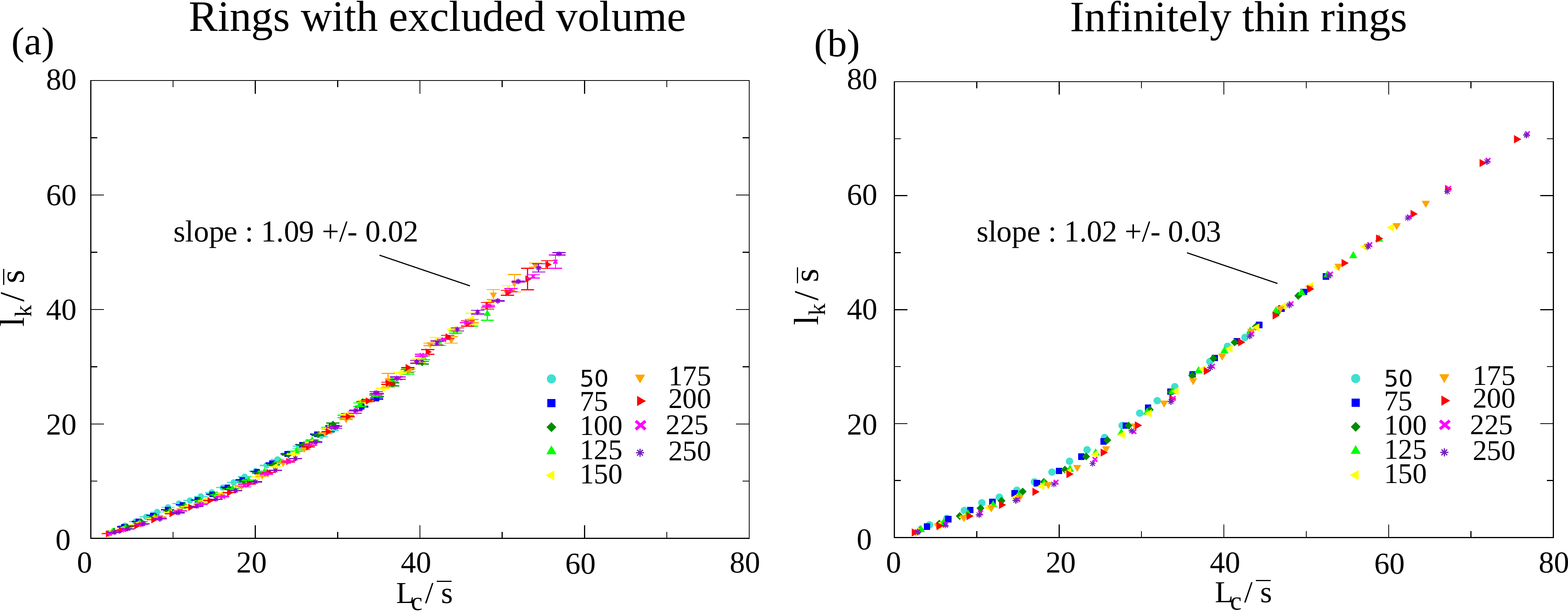}
\caption{Collapse of the data points shown in figure 
  \ref{fig:supp_mat_4}b (rings with excluded volume, as in main paper) and
  \ref{fig:supp_mat_4}d (rings without excluded volume) obtained by
  rescaling $l_k$ and $L_c$ by the deflection length, $\bar{s}$. The same
  deflection length as in eqn. 3 of the main paper was used in both cases.
  Note that the slope of the line fitting the data for $L_c/\bar{s}>40$ is
  compatible between the two cases.}
\end{figure}

\begin{figure}[h!]
\includegraphics[width=0.8\textwidth]{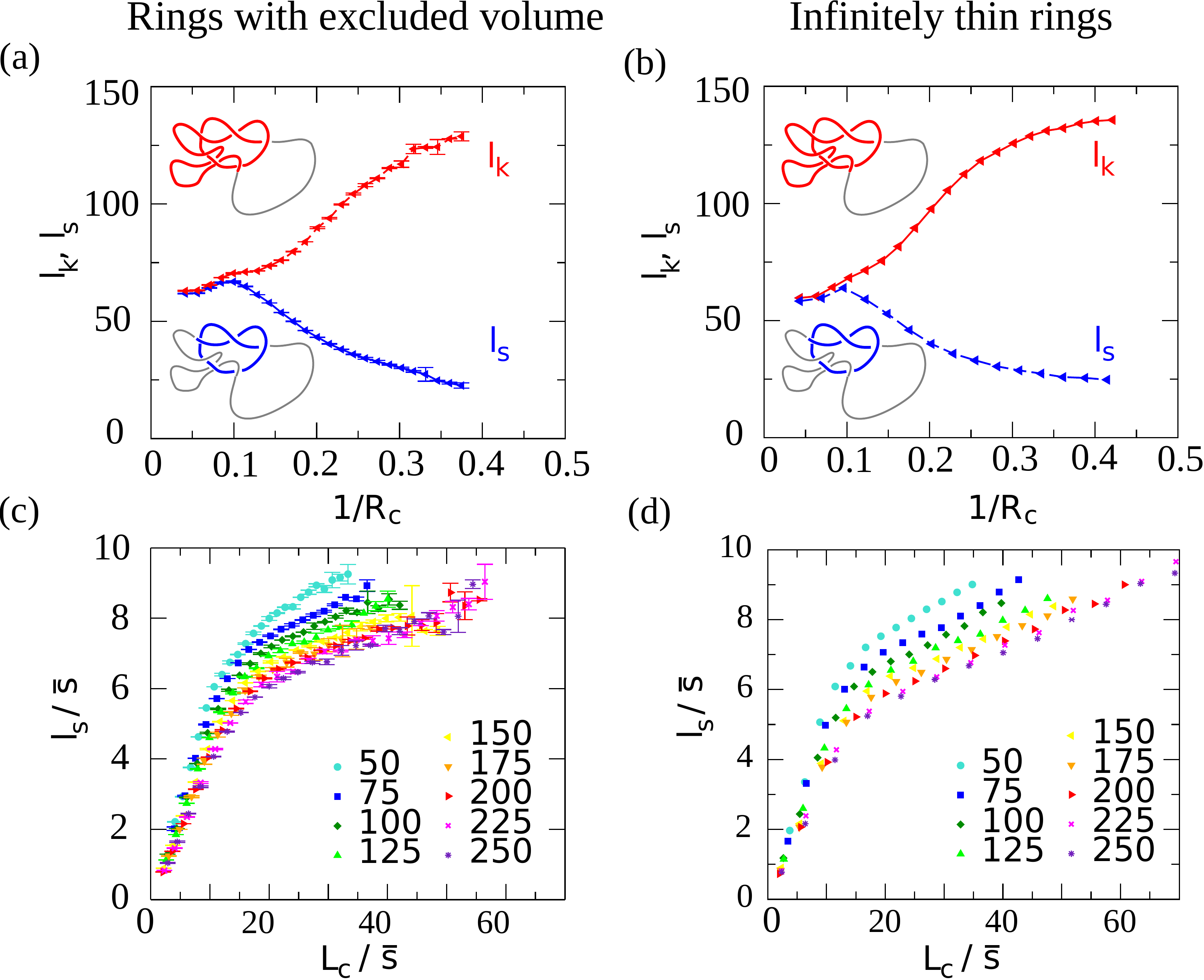}
\caption{Top: dependence of knot length, $l_k$, and the shortest knot
  length, $l_s$, on confinement for rings of $N = 150$ segments with excluded volume (panel a) and without excluded volume, (panel b). Bottom:
  scatter plots of $l_s/\bar{s}$ versus $L_c/\bar{s}$ for rings with excluded volume (panel c) and without excluded volume (panel d). }
\end{figure}

\end{document}